\documentclass{IOS-Book-Article}
\usepackage{graphicx}
\usepackage{mathptmx}
\usepackage{soul}\setuldepth{article}
\usepackage{caption}
\usepackage{graphicx}
\usepackage{wrapfig}
\usepackage{lscape}
\usepackage{rotating}
\usepackage{epstopdf}

\def\hb{\hbox to 11.5 cm{}}

\begin{document}

\pagestyle{headings}
\def\thepage{}

\begin{frontmatter}              

\title{Understanding EEG signals for subject-wise Definition of Armoni Activities}

\markboth{}{\hb}

\author[A]{\fnms{Kislay} \snm{Raj}}
\thanks{Corresponding Author:Aditya Singh, DEIM, URV Spain; E-mail: aditya.singh@urv.cat},
\author[B]{\fnms{Aditya} \snm{Singh}}, %
\author[A]{\fnms{Abhishek} \snm{Mandal}}, %
\author[A]{\fnms{Teerath} \snm{Kumar}}, %
\author[C]{\fnms{Arunabha M.} \snm{  Roy}} %

\address[A]{School of Computing, Faculty of Engineering and Computing, Dublin City University, Dublin, Ireland}
\address[B]{Instituto de Robotica para la Dependencia, Sitges, Spain}
\address[C]{Aerospace Engineering Department, University of Michigan, Ann Arbor, MI
48109, USA}
\begin{abstract}
In a growing world of technology, psychological disorders became a challenge to be solved. The methods used for cognitive stimulation are very conventional and based on one-way communication, which only rely on the material or method used for training of an individual. It doesn't use any kind of feedback from the individual to analyze the progress of the training process. We have proposed a closed-loop methodology to improve the cognitive state of a person with ID (Intellectual disability). We have used a platform named ‘Armoni’, for providing training to the intellectually disabled individuals. The learning is performed in a closed-loop by using feedback in the form of change in affective state. For feedback to the Armoni, an EEG (Electroencephalograph) headband is used. All the changes in EEG are observed and classified against the change in the mean and standard deviation value of all frequency bands of signal. This comparison is being helpful in defining every activity with respect to change in brain signals. In this paper, we have discussed the process of treatment of EEG signal and its definition against the different activities of Armoni. We have tested it on 6 different systems with different age groups and cognitive levels. 
\end{abstract}
\begin{keyword}
Electroencephalography\sep Armoni \sep Cognitive Stimulation\sep Intellectual Disability 
\end{keyword}
\end{frontmatter}
\markboth{\hb}{\hb}
\section{Introduction}
Deep learning (DL) has been very successful in a range of different data domains including image classification 
\cite{aleem2022random,kumar2021class,khan2022introducing,ranjbarzadeh2022brain, haseli2022hecon,  kumar2021binary, kumarforged,roy2022computer, tataei2023glioma,jamil2022distinguishing,deng2009imagenet,ranjbarzadeh2022brain,tataei2022glioma}, audio~\cite{kumar2020intra, chandio2021audd, park2020search, turab2022investigating}, text~\cite{ullah2021rweetminer,kowsari2019text}, 
 computer vision \cite{chandio2022precise,roy2022wildect,singh2023deep}, object detection \cite{roy2021deep,roy2022fast,roy2022real}, object segmentation~\cite{ranjbarzadeh2022deep,ranjbarzadeh2022breast,ranjbarzadeh2022brain,ranjbarzadeh2022mrfe},brain-computer interface \cite{roy2022efficient,roy2022adaptive,roy2022multi}, 
and across diverse scientific disciplines \cite{bose2022accurate,khan2022sql}.
Intellectual Disability (ID) with DL has rarely been discussed and ID became a rising problem in modern society. Its negligence can result in the form of psychological disorders and symptoms. A person with ID (Intellectual Disability) faces many issues while behaving socially. The sensory response to an impromptu situation became so long and it suffers to a huge lag. Some methods \cite{r1,r2} are used for training of persons suffering from psychological disorders. These methods include clinical therapies \cite{r3}, personal assistance \cite{r4} and Intelligent System based solutions. These systems need a proper guidance from a psychological expert to conduct any treatment activity. 

In our work, we have used an 'Armoni' \cite{r5} platform for cognitive stimulation of person with ID. It features 15 activities for training of person with ID. These all activities are developed under the supervision of psychologists and can be customized as per the need. It measures the intellectual performance of any person in total eleven cognitive areas and generates a report for professionals and family. This system is very useful for training of persons with intellectual disability. It provides a visual-audio interface for training and offers a touch screen to provide haptic interface for activity access. Every activity of Armoni is defined in several predictors as: Comprehension, Visuo perception, Long Term Memory, Visuo construction, Comprehension, Naming Ability, Visual Memory, and Verbal Memory. Every activity can be associated with one or more than predictions. 

The Armoni is initially used as a training platform for cognitive purposes. Its application is only defined in one-way. The training process relies on the potential of the activities in stimulation effectively. In case of ID, the effect of any training differs due to varying perception ability of each individual. The potential of treatment process can be questioned against the effect of a particular activity. It requires a feedback of the stimuli response for every activity, which can be used as a parameter to evaluate the effect of training.

\begin{figure}[!ht]
    \centering
    \includegraphics[scale=0.68]{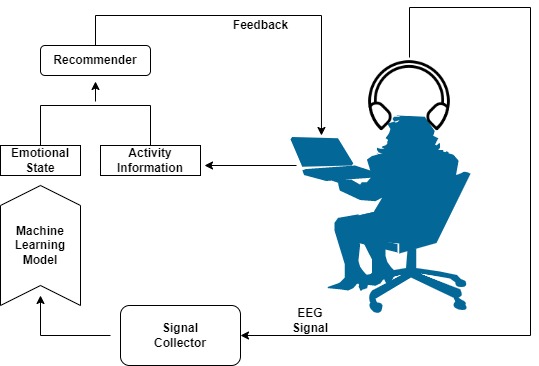}
    \caption{Schematic Diagram of Proposed Methodology. It uses a tablet based training platform for cognitive stimulation. EEG headband is used for measuring brain signals. Signals are then filtered and classified to different dimension of emotions, which is used for generating feedback.}
    \label{fig_example}
\end{figure}

We have proposed a human in the loop method for making training effective. In the proposed method, the reflections in brain neurons by a particular training activity is being recorded, analyzed and used as feedback to improve the training process. We have used the brain signals (recorded in the form of EEG) for the detection of engagement of brain in the individual. An analytical model is used to classify the input signals into the perception and impact on brain. This analysis will be used as a feedback to the armoni system to host the next activity for training. The feedback recommendation is generated by a dedicated recommendation system. The overview of the proposed closed-loop model is shown in Figure \ref{fig_example}.

The stated process for rehabilitation involves several processes like observation, recommendation, and training. It is being done is several steps as shown below:
\begin{itemize}
    \item An Individual will undergo a training process by using touch screen of Armoni.
    \item An EEG device will measure changes in brain signals.
    \item Signals will feed to a pre-trained analytical model. The model is trained on standard dataset made by previous observations for signals versus stimuli changes.
    \item Model will predict the change in stimuli corresponding to the changes in brain signal of the participant.
    \item Changes predicted by the model and current activity parameters (activity name, response time of user, difficulty level) will be compared for the improvement of learning procedure.
    \item Comparison results will be fed to the Armoni Application for change in training procedure.
\end{itemize}

In this paper, we have discussed the method of data collection, its processing, and relation of EEG signal with respect to different parts of brain. We have converted the time domain EEG signal to frequency domain signal and break it into different frequency bands (Alpha, beta, Gamma, Delta, Theta) as per the range of values. It uses four signals (TP9, AF7, AF8, TP10) in which two are from left hemisphere and two are from right hemisphere. Each activity and subject is examined in terms for dominant hemisphere, change occurred due to activity with respect to normal condition. Each hemisphere and electrode is compared to define each analysis. 

This paper will address the method used till the process of signal processing and categorization in the process of Cognitive Stimulation. It propose a detailed explanation of EEG signals into its different components. The rest of the article is arranged as: Section 2 discusses the methodology for the proposed method. In section 3, results for emotion classification are discussed. Section 4 concludes the problem statement and discusses the future scope of the problem.

\section{Previous work on Emotion Understanding using EEG}
Sumethee et al \cite{r8} have proposed a review focusing on principles and applications of EEG in food research. It records the EEG signal from the scalp of different subjects and noted down the changes occurred in the signal by smell, taste and look of different food items. This survey is then used to make changes in product packaging and its presentation in the market. In \cite{r9}, a short-time emotion recognition technique is used. It boycotts the use of handcrafted features and proposed a spiking neural network (SNN) to detect the changes in EEG signals. This method is tested on DEAP and MAHNOB-HCI dataset and claims an accuracy of 78.9\% and 79.4\% for arousal detection. In \cite{r10}, a review of current trends in EEG based emotion recognition is discussed. In \cite{r11}, EEG signal is used to predict emotional state in terms of change in Arousal and Valence and cortisol content from the saliva is used as a reliable measuring parameter for emotion change. It suggest the effectiveness of valence measurement in prediction of pleasantness. In \cite{r12}, EEG features like Hjorth features are tested on a cross-subject scenario. In \cite{r13}, the EEG signal is merged with the voice input to make a multi-modal approach to solve emotion recognition. 

In every work, the use of EEG differs on the basis of application. In our case, the subjects are with ID or normal conditions both. We have adapted a left and right hemisphere based analysis to distinguish the mind activities in terms of holistic and analytic approach. 

\section{Proposed Methodology}
This section describes the problem definition for solving cognitive stimulation. It contains collection of brain signals. The breaking of every component in different frequency components and its impact is discussed. It also discusses the brain structure and the impact of dominance of every part of brain.  

\subsection{Problem Definition}
The training of persons with ID is a tough task because of the absence of reliable feedback for its condition. The EEG signals is used for detecting the change in the mind. The EEG signal reflects the change in mind conditions in the form of change of electrical (voltage) potential. We have measured signal in terms of four electrodes: AF7, AF8 from forehead and TP9, TP10 from ear. 

In our problem, the EEG signal is used for analyzing feedback from the user. Signal is converted and processed in frequency domain and it is broken to different frequency bands. These components are: 
\begin{itemize}
    \item Delta: 0-4 Hz
    \item Theta: 4-8 Hz
    \item Alpha: 8-12 Hz
    \item Beta: 12-30 Hz
    \item Gamma: 30-45 Hz
    
\end{itemize}

After having frequency band component for every signal from different electrodes, it is necessary to classify it towards the dominance of the hemisphere. Every activity on Armoni is being defined on the basis of hemisphere dominance and Activity Dominance with respect to talking activity performed by every subject. 

\subsection{Brain Signal Reading}
\begin{figure}[!ht]
    \centering
    \includegraphics[scale=0.68]{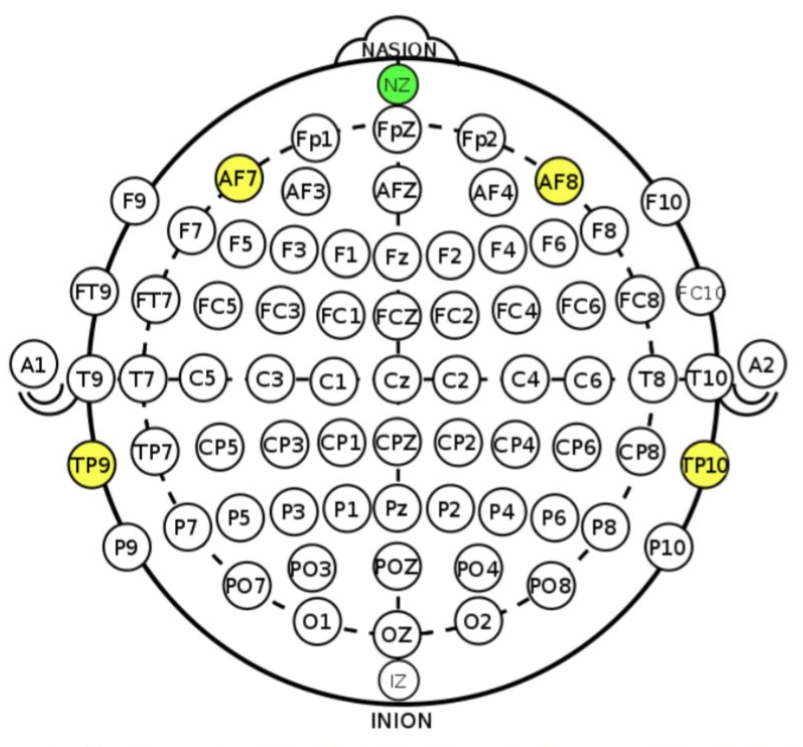}
    \caption{Placement of sensor electrodes. Two electrodes on left Hemisphere: TP9, AF7. Two electrodes on right hemisphere: TP10, AF8. One neutral or reference electrode on the center of the forehead. AF7 and AF8 are placed on the forehead and TP9 and TP10 are placed behind the ears of the participant.}
    \label{Brain}
\end{figure}

A MUSE EEG headband \cite{r6} is used for capturing brain signals. It considers 5 electrodes for reading brain signals. As shown in Figure \ref{Brain}, four brain locations (AF7, AF8, TP9, TP10) are used to capture signals and one neutral point (NZ) is used as reference for other electrodes. Out of four, two electrodes (AF7, AF8) are placed on forehead and other two (TP9, TP10) are placed behind the both ears. The EEG signals are classified as per left and right hemisphere.

Banich and Amp \cite{r7} have proposed a wonderful definition for human mind by classifying it to left and right hemisphere. It considers the left and right brain as tree's and forest. This concept depicts the nature of both hemispheres. The tree nature means the objectivity in the selection of task and forest means to treat any problem as a whole. It means that the left brain focus more on the reason for every move and right brain considers the essence of the whole problem. In our case, we have considered left and right hemisphere electrodes as a separate combination.

\subsection{Converting Time Signal to Frequency Signal}
\begin{figure}[!ht]
    \centering
    \includegraphics[scale=0.44]{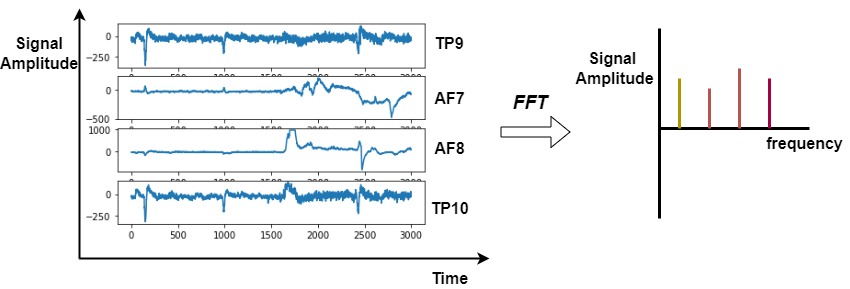}
    \caption{Time Domain to frequency domain conversion for categorization based analysis.}
    \label{time-freq}
\end{figure}

The processing of EEG signals in time domain is tough and relevant features are not available in time domain. The signal is being transformed to frequency domain by using Fourier transform. As shown in Figure \ref{time-freq}, the time domain signal from all four electrodes is being converted to a frequency domain representation. 

\begin{equation}
    frequency\: domain = fft(time\: domain)
\end{equation}

The frequency domain value for each signal is being partitioned to corresponding frequency bands. Every frequency band represent to a different degree of entropy in the mind. The 'Delta' component represents the most pleasant condition of the mind like sleeping and deep meditation and 'Gamma' component represent the most dynamics condition of the mind. Using this nature of different frequency components with tree and forest representation will result in the form of a complete analysis. 

\subsection{Evaluation of changes observed in EEG signals}
After calculating band wise information, it needs to analyze that which hemisphere is more attentive during training. We have considered Gamma wave as a reference out of five frequency bands because it is more associated with the quick decisions and training activities. The signals associated with the left and right hemisphere are added for comparison. 

\begin{equation}
    sum_{left\_hemisphere} =Change\_in(\gamma_{AF7} + \gamma_{TP9})
\end{equation}
\begin{equation}
    sum_{right\_hemisphere} - Change\_in(\gamma_{AF8} + \gamma_{TP10})
\end{equation}

The comparison of gamma waves will depict the dominant hemisphere during any training activity. This hemisphere selection will leads to calculate the effect of activity. The effect can be in the form of increase and decrease in $\gamma$ component.

\begin{equation}
    \gamma_{change} = \gamma_{training}-\gamma_{talking}
\end{equation}

If the change is positive than training activity dominates the mind activity and if change is negative than training activity does not dominate. The dominance in activities and hemisphere will help in deciding the next activity used for training.

\section{Results}
We have recorded the EEG data on various subjects while training on different activities from Armoni. It is tested on three Activities (Puzzle solving, memory, Paint or object Rain) for every subject. Every activity has a specific nature in the form of predictors, which ensembles that in which way it is going to effect the subject. 

Firstly, The subject undergoes through a talking process. The talking process data is used as reference to other training activities. Every activity of training create changes in brain signals which are compared in terms of the gamma component of every hemisphere.

\begin{sidewaysfigure}[ht]
    \includegraphics[scale=0.54]{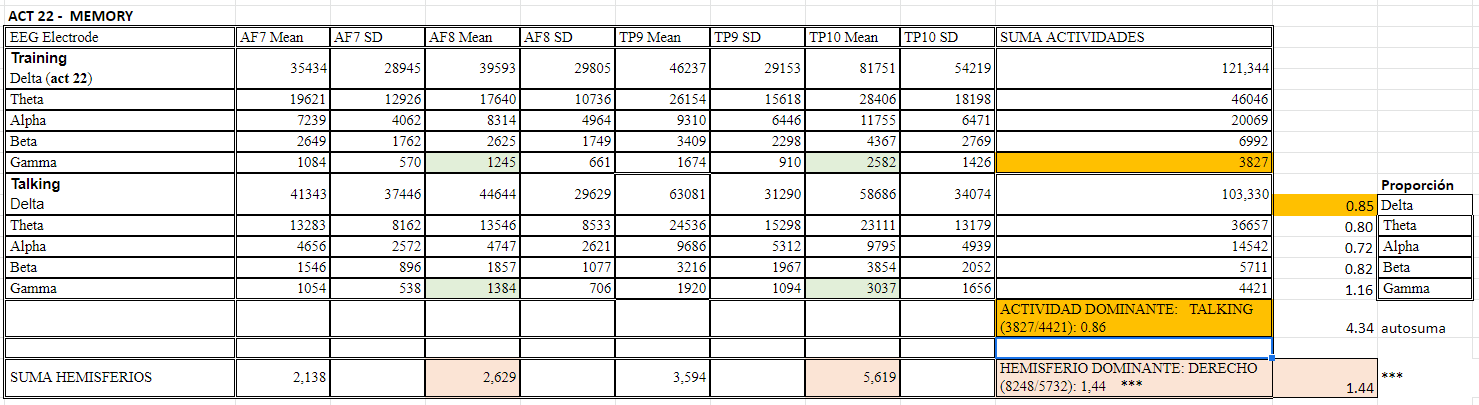}
    \caption{Report consisting the changes in different signals. }
    \label{report}
\end{sidewaysfigure}

In Figure \ref{report}, a report containing the comparison for training and talking activities is shown. The gamma waves are added first hemisphere wise and compared to find the dominant hemisphere and after finding the dominant hemisphere, $\gamma$ waves are compared for both activities for that particular hemisphere. 

In Table \ref{tab1}, we have compared the result for two different activities for 6 different subjects. The activities used in table are Puzzle and Memory. These activities have different cognitive dimensions for training. 

\begin{table}
\caption{\centering 
Subject-wise effect of Memory and Puzzle activities of Armoni
\textit{IVM: Instant Verbal Memory, VP: Visuoperception, VC: Visuoconstruction, C: Comprehension}}
\textit{Right: Holistic Approach, Left: Analytic Approach}
\label{ranjit_1l}
\begin{tabular}{|l|l|l|l|l|}
\hline
Subject & Activity & Cognitive Domains & Dominant Hemisphere & Dominant Activity\\
\hline

Jaume & Memory & IVM, VP& Left& Training\\
Sara & Memory &IVM, VP& Left& Talking \\
Oscar & Memory &IVM, VP& Right& Talking\\
Yolanda & Memory &IVM, VP& Right& Training\\
Gaurav &  Memory &IVM, VP& Left& Training\\
Benigno & Memory &IVM, VP& Left& Training\\
\hline
Jaume & Puzzle & VP, VC, C & Left& Training\\
Sara & Puzzle & VP, VC, C & Right& Training  \\
Oscar & Puzzle & VP, VC, C & Right& Talking\\
Yolanda & Puzzle & VP, VC, C & Left& Talking\\
Gaurav &  Puzzle & VP, VC, C &Left &Training\\

\hline

\end{tabular}
\label{tab1}
\end{table}

\section{Conclusion}
We have presented a method for analyzing the effect of training activities used in Armoni. The analysis of people with ID is presented in terms of the affect of the activity on mind and the involvement of mind. This study is used in the selection of further activities for training of person with ID. This work is the first known work in our knowledge for using EEG for online training analysis and improvement. We have tested it with 6 subjects with varying mental abilities on different activities of Armoni.  In future, this method can be extended with more specific description of the brain activities and by using a autonomous process of training.

\section{Acknowledgement}
This work is done in a collaboration between Instituto de Robotica, IIIT Allahabad, and URV Spain. All the facilities are provided in the campus of Instituto de Robotica para la Dependencia, Sitges. 

\section{Conflict of Interest}
The authors declare that there is no financial interests or personal relationships that could have appeared to influence the work reported in this paper. 

\section{Data Availability}
The data associated with current research is available upon reasonable request.

\end{document}